%%                              JNL.TEX
%%
%%                This is JNL.TEX Version 0.3 as of 6/12/85.
%%
%%      This is a set of TeX 82 macros designed to produce scientific
%%      papers with a minimum of fuss and using as much of plain.tex as
%%      possible.  The user need only know what is in the TeXbook, and
%%      the macros under ``user definitions'' below.  Also, the user
%%      definitions are intended to be as simple as possible, so that
%%      the user may change them as desired.

%%
%%  Font definitions suitable for the IMAGEN (Written by Tony Kennedy)
%%

%  Define a whole menagerie of pseudo-12pt fonts

\font\twelverm=cmr10 scaled 1200    \font\twelvei=cmmi10 scaled 1200
\font\twelvesy=cmsy10 scaled 1200   \font\twelveex=cmex10 scaled 1200
\font\twelvebf=cmbx10 scaled 1200   \font\twelvesl=cmsl10 scaled 1200
\font\twelvett=cmtt10 scaled 1200   \font\twelveit=cmti10 scaled 1200

\skewchar\twelvei='177   \skewchar\twelvesy='60

%  Define \...point macros to change fonts and spacings consistently

\def\twelvepoint{\normalbaselineskip=12.4pt
  \abovedisplayskip 12.4pt plus 3pt minus 9pt
  \belowdisplayskip 12.4pt plus 3pt minus 9pt
  \abovedisplayshortskip 0pt plus 3pt
  \belowdisplayshortskip 7.2pt plus 3pt minus 4pt
  \smallskipamount=3.6pt plus1.2pt minus1.2pt
  \medskipamount=7.2pt plus2.4pt minus2.4pt
  \bigskipamount=14.4pt plus4.8pt minus4.8pt
  \def\rm{\fam0\twelverm}          \def\it{\fam\itfam\twelveit}%
  \def\sl{\fam\slfam\twelvesl}     \def\bf{\fam\bffam\twelvebf}%
  \def\mit{\fam 1}                 \def\cal{\fam 2}%
  \def\tt{\twelvett}
  \textfont0=\twelverm   \scriptfont0=\tenrm   \scriptscriptfont0=\sevenrm
  \textfont1=\twelvei    \scriptfont1=\teni    \scriptscriptfont1=\seveni
  \textfont2=\twelvesy   \scriptfont2=\tensy   \scriptscriptfont2=\sevensy
  \textfont3=\twelveex   \scriptfont3=\twelveex  \scriptscriptfont3=\twelveex
  \textfont\itfam=\twelveit
  \textfont\slfam=\twelvesl
  \textfont\bffam=\twelvebf \scriptfont\bffam=\tenbf
  \scriptscriptfont\bffam=\sevenbf
  \normalbaselines\rm}

%       tenpoint

%%
%%      Various internal macros
%%

\def\beginlinemode{\endmode
  \begingroup\parskip=0pt \obeylines\def\\{\par}\def\endmode{\par\endgroup}}
\def\beginparmode{\endmode
  \begingroup \def\endmode{\par\endgroup}}
\let\endmode=\par
{\obeylines\gdef\
{}}
\def\singlespace{\baselineskip=\normalbaselineskip}

\def\oneandahalfspace{\baselineskip=\normalbaselineskip
  \multiply\baselineskip by 3 \divide\baselineskip by 2}
\def\doublespace{\baselineskip=\normalbaselineskip \multiply\baselineskip by 2}

\newcount\firstpageno
\firstpageno=2
\footline={\ifnum\pageno<\firstpageno{\hfil}%
\else{\hfil\twelverm\folio\hfil}\fi}
\let\rawfootnote=\footnote              % We must set the footnote style
\def\footnote#1#2{{\rm\singlespace\parindent=0pt\rawfootnote{#1}{#2}}}
\def\raggedcenter{\leftskip=4em plus 12em \rightskip=\leftskip
  \parindent=0pt \parfillskip=0pt \spaceskip=.3333em \xspaceskip=.5em
  \pretolerance=9999 \tolerance=9999
  \hyphenpenalty=9999 \exhyphenpenalty=9999 }
\def\dateline{\rightline{\ifcase\month\or
  January\or February\or March\or April\or May\or June\or
  July\or August\or September\or October\or November\or December\fi
  \space\number\year}}
\def\received{\vskip 3pt plus 0.2fill
 \centerline{\sl (Received\space\ifcase\month\or
  January\or February\or March\or April\or May\or June\or
  July\or August\or September\or October\or November\or December\fi
  \qquad, \number\year)}}

%%
%%      Page layout, margins, font and spacing (feel free to change)
%%

\hsize=6.5truein
%\hoffset=1truein
\vsize=8.9truein
%\voffset=1truein
\parskip=\medskipamount
\twelvepoint            % selects twelvepoint fonts (cf. \tenpoint)
\doublespace            % selects double spacing for main part of paper (cf.
                        %       \singlespace, \oneandahalfspace)
\overfullrule=0pt       % delete the nasty little black boxes for overfull box

%%
%%      The user definitions for major parts of a paper (feel free to change)
%%

    % Preprint number at upper right of title page

\def\title                      %  Title on title page
  {\null\vskip 3pt plus 0.2fill
   \beginlinemode \doublespace \raggedcenter \bf}

\def\author                     %  Author(s) name(s)  on title page
  {\vskip 3pt plus 0.2fill \beginlinemode
   \singlespace \raggedcenter}

\def\affil                      % Affiliations (can intermix with \author)
  {\vskip 3pt plus 0.1fill \beginlinemode
   \oneandahalfspace \raggedcenter \sl}

\def\abstract                   % Begin abstract
  {\vskip 3pt plus 0.3fill \beginparmode
   \doublespace \narrower ABSTRACT: }

\def\endtitlepage               % End title page, begin body of paper
  {\endpage                     %       This subsumes \body
   \body}

\def\body                       % Begin text body;  can be used to end
  {\beginparmode}               % \title, \author, \affil, \abstract,
                                % \reference, or \figurecaption modes

\def\subhead#1{                 % Subhead;  NOTE enclose the text in {}
  \vskip 0.25truein             % e.g., \subhead{A. History of the Problem}
  {\raggedcenter #1 \par}
   \nobreak\vskip 0.25truein\nobreak}

\def\refto#1{$[{#1}]$}           % For references in text as superscript

\def\references                 % Begin references -- basic format is Phys Rev
  {\subhead{References}         % I.e., volume, page, year (space after commas)
   \beginparmode
   \frenchspacing \parindent=0pt \leftskip=1truecm
   \parskip=8pt plus 3pt \everypar{\hangindent=\parindent}}

\gdef\refis#1{\indent\hbox to 0pt{\hss#1.~}}    % Ref list numbers.

\gdef\journal#1, #2, #3, 1#4#5#6{               % Journal reference.  Comma set
    {\sl #1~}{\bf #2}, #3, (1#4#5#6)}           % off: name, vol, page, year

\def\refstylenp{                % Nucl Phys(or Phys Lett) ref style: V, Y, P
  \gdef\refto##1{ [##1]}                                % Reference in text []
  \gdef\refis##1{\indent\hbox to 0pt{\hss##1)~}}        % Ref list numbers)
  \gdef\journal##1, ##2, ##3, ##4 {                     % Journal reference
     {\sl ##1~}{\bf ##2~}(##3) ##4 }}

\def\refstyleprnp{              % Input like pr, output like np!!
  \gdef\refto##1{ [##1]}                                % Reference in text []
  \gdef\refis##1{\indent\hbox to 0pt{\hss##1)~}}        % Ref list numbers)
  \gdef\journal##1, ##2, ##3, 1##4##5##6{               % Journal reference
    {\sl ##1~}{\bf ##2~}(1##4##5##6) ##3}}

\def\endreferences{\body}

\def\figurecaptions             % Begin figure captions
  { \beginparmode
   \subhead{Figure Captions}
}

\def\endpage                    %  Eject a page
  {\vfill\eject}

\def\endpaper                   %  Ways to say goodbye
  {\endmode\vfill\supereject}

%%
%%      Various little user definitions
%%

\def\ref#1{Ref. #1}                     %       for inline references
\def\Ref#1{Ref. #1}                     %       ditto

          % For citation of equation numbers
        %       ditto
                     %       ditto
                     %       ditto
                   %       ditto
                   %       ditto
\def\frac#1#2{{\textstyle{#1 \over #2}}}

\def\sla{\raise.15ex\hbox{$/$}\kern-.57em}
\def\leaderfill{\leaders\hbox to 1em{\hss.\hss}\hfill}
\def\twiddle{\lower.9ex\rlap{$\kern-.1em\scriptstyle\sim$}}
\def\bigtwiddle{\lower1.ex\rlap{$\sim$}}
\def\gtwid{\mathrel{\raise.3ex\hbox{$>$\kern-.75em\lower1ex\hbox{$\sim$}}}}
\def\ltwid{\mathrel{\raise.3ex\hbox{$<$\kern-.75em\lower1ex\hbox{$\sim$}}}}
\def\square{\kern1pt\vbox{\hrule height 1.2pt\hbox{\vrule width 1.2pt\hskip 3pt
   \vbox{\vskip 6pt}\hskip 3pt\vrule width 0.6pt}\hrule height 0.6pt}\kern1pt}

\def\m@th{\mathsurround=0pt }
\def\leftrightarrowfill{$\m@th \mathord\leftarrow \mkern-6mu
 \cleaders\hbox{$\mkern-2mu \mathord- \mkern-2mu$}\hfill
 \mkern-6mu \mathord\rightarrow$}
\def\overleftrightarrow#1{\vbox{\ialign{##\crcr
     \leftrightarrowfill\crcr\noalign{\kern-1pt\nointerlineskip}
     $\hfil\displaystyle{#1}\hfil$\crcr}}}

%% *********** New stuff follows *******************

\font\titlefont=cmr10 scaled\magstep3

\def\martinstyletitle                      %  Title on title page
  {\null\vskip 3pt plus 0.2fill
   \beginlinemode \doublespace \raggedcenter \titlefont}

\font\twelvesc=cmcsc10 scaled 1200

\def\author                     %  Author(s) name(s)  on title page
  {\vskip 3pt plus 0.2fill \beginlinemode
   \singlespace \raggedcenter\twelvesc}

%%
%%      AmSTeX compatability definitions
%%
%%      To run a TeX file originally intended for AmSTeX, only small changes
%%      should be necessary (I hope).  Use the line \input jnl at the start.
%%      Remove the lines \input amstex, \documentstyle{itpjnl} at the
%%      beginning;  also remove all the page layout stuff (\parindent=1cm,
%%      \hsize=5.28125in etc.)  The page layout is now done automatically.
%%      Also OMIT the qualifier \magnification=1200 when you IMPRINT the
%%      .dvi file.  (\TagsOnRight is harmless, you can take it out or leave
%%      it in.)  I believe most AmSTeX will work with no change.  One problem
%%      is \footnote, which is a little different in that it now needs to
%%      have an explicit asterisk *  (or whatever) included, like this:
%%              \footnote*{Text winds up at bottom of page.}
%%      This is discussed on p. 116 of the TeXbook.  IGNORE the AmSTeX
%%      documentation (if you can call it that);  refer to the TeXbook.
%%
%%      Note that many commands in AmSTeX have their equivalents in the
%%      TeXbook, perhaps with different names and slightly differing
%%      usage. E.g., the old \align in AmSTeX is replaced by \eqalign
%%      (p. 190) and \aligntag is replaced by \eqalignno (p. 192).
%%      \align and \aligntag still work, but I recommend that you use
%%      \eqalign and \eqalignno in documents run under jnl.
%%
%%      See me if you have any problems  -- Doug.
%%

\def\heading                            % Heading
  {\vskip 0.5truein plus 0.1truein      % e.g., \heading I. NOTES \endheading
   \beginparmode \def\\{\par} \parskip=0pt \singlespace \raggedcenter}

\def\subheading                         % Subheading
  {\vskip 0.25truein plus 0.1truein     % e.g., \subheading{A. The Problem}
   \beginlinemode \singlespace \parskip=0pt \def\\{\par}\raggedcenter}

\def\tag#1$${\eqno(#1)$$}

\def\align#1$${\eqalign{#1}$$}

\def\aligntag#1$${\gdef\tag##1\\{&(##1)\cr}\eqalignno{#1\\}$$
  \gdef\tag##1$${\eqno(##1)$$}}

\def\endaligntag{}

\def\overset #1\to#2{{\mathop{#2}\limits^{#1}}}
\def\underset#1\to#2{{\let\next=#1\mathpalette\undersetpalette#2}}
\def\undersetpalette#1#2{\vtop{\baselineskip0pt
\ialign{$\mathsurround=0pt #1\hfil##\hfil$\crcr#2\crcr\next\crcr}}}

%%
%%      Various little user definitions
%%

\def\ref#1{Ref.~#1}                     %       for inline references
\def\Ref#1{Ref.~#1}                     %       ditto
\def\[#1]{[\cite{#1}]}
\def\cite#1{{#1}}
\def\(#1){(\call{#1})}
\def\call#1{{#1}}
\def\taghead#1{}
\def\frac#1#2{{#1 \over #2}}

\def\12{{1\over2}}

\def\sla{\raise.15ex\hbox{$/$}\kern-.57em}
\def\leaderfill{\leaders\hbox to 1em{\hss.\hss}\hfill}
\def\twiddle{\lower.9ex\rlap{$\kern-.1em\scriptstyle\sim$}}
\def\bigtwiddle{\lower1.ex\rlap{$\sim$}}
\def\gtwid{\mathrel{\raise.3ex\hbox{$>$\kern-.75em\lower1ex\hbox{$\sim$}}}}
\def\ltwid{\mathrel{\raise.3ex\hbox{$<$\kern-.75em\lower1ex\hbox{$\sim$}}}}
\def\square{\kern1pt\vbox{\hrule height 1.2pt\hbox{\vrule width 1.2pt\hskip 3pt
   \vbox{\vskip 6pt}\hskip 3pt\vrule width 0.6pt}\hrule height 0.6pt}\kern1pt}
\def\tdot#1{\mathord{\mathop{#1}\limits^{\kern2pt\ldots}}}

\def\pmb#1{\setbox0=\hbox{#1}%
  \kern-.025em\copy0\kern-\wd0
  \kern  .05em\copy0\kern-\wd0
  \kern-.025em\raise.0433em\box0 }

\catcode`@=11
\newcount\tagnumber\tagnumber=0

\immediate\newwrite\eqnfile
\newif\if@qnfile\@qnfilefalse
\def\write@qn#1{}
\def\writenew@qn#1{}
\def\w@rnwrite#1{\write@qn{#1}\message{#1}}
\def\@rrwrite#1{\write@qn{#1}\errmessage{#1}}

\def\taghead#1{\gdef\t@ghead{#1}\global\tagnumber=0}
\def\t@ghead{}

\expandafter\def\csname @qnnum-3\endcsname
  {{\t@ghead\advance\tagnumber by -3\relax\number\tagnumber}}
\expandafter\def\csname @qnnum-2\endcsname
  {{\t@ghead\advance\tagnumber by -2\relax\number\tagnumber}}
\expandafter\def\csname @qnnum-1\endcsname
  {{\t@ghead\advance\tagnumber by -1\relax\number\tagnumber}}
\expandafter\def\csname @qnnum0\endcsname
  {\t@ghead\number\tagnumber}
\expandafter\def\csname @qnnum+1\endcsname
  {{\t@ghead\advance\tagnumber by 1\relax\number\tagnumber}}
\expandafter\def\csname @qnnum+2\endcsname
  {{\t@ghead\advance\tagnumber by 2\relax\number\tagnumber}}
\expandafter\def\csname @qnnum+3\endcsname
  {{\t@ghead\advance\tagnumber by 3\relax\number\tagnumber}}

\def\equationfile{%
  \@qnfiletrue\immediate\openout\eqnfile=\jobname.eqn%
  \def\write@qn##1{\if@qnfile\immediate\write\eqnfile{##1}\fi}
  \def\writenew@qn##1{\if@qnfile\immediate\write\eqnfile
    {\noexpand\tag{##1} = (\t@ghead\number\tagnumber)}\fi}
}

\def\callall#1{\xdef#1##1{#1{\noexpand\call{##1}}}}
\def\call#1{\each@rg\callr@nge{#1}}

\def\each@rg#1#2{{\let\thecsname=#1\expandafter\first@rg#2,\end,}}
\def\first@rg#1,{\thecsname{#1}\apply@rg}
\def\apply@rg#1,{\ifx\end#1\let\next=\relax%
\else,\thecsname{#1}\let\next=\apply@rg\fi\next}

\def\callr@nge#1{\calldor@nge#1-\end-}
\def\callr@ngeat#1\end-{#1}
\def\calldor@nge#1-#2-{\ifx\end#2\@qneatspace#1 %
  \else\calll@@p{#1}{#2}\callr@ngeat\fi}
\def\calll@@p#1#2{\ifnum#1>#2{\@rrwrite{Equation range #1-#2\space is bad.}
\errhelp{If you call a series of equations by the notation M-N, then M and
N must be integers, and N must be greater than or equal to M.}}\else %
{\count0=#1\count1=#2\advance\count1 by1\relax\expandafter\@qncall\the\count0,%
  \loop\advance\count0 by1\relax%
    \ifnum\count0<\count1,\expandafter\@qncall\the\count0,%
  \repeat}\fi}

\def\@qneatspace#1#2 {\@qncall#1#2,}
\def\@qncall#1,{\ifunc@lled{#1}{\def\next{#1}\ifx\next\empty\else
  \w@rnwrite{Equation number \noexpand\(>>#1<<) has not been defined yet.}
  >>#1<<\fi}\else\csname @qnnum#1\endcsname\fi}

\let\eqnono=\eqno
\def\eqno(#1){\tag#1}
\def\tag#1$${\eqnono(\displayt@g#1 )$$}

\def\aligntag#1\endaligntag
  $${\gdef\tag##1\\{&(##1 )\cr}\eqalignno{#1\\}$$
  \gdef\tag##1$${\eqnono(\displayt@g##1 )$$}}

\def\eqalignno#1{\displ@y \tabskip\centering
  \halign to\displaywidth{\hfil$\displaystyle{##}$\tabskip\z@skip
    &$\displaystyle{{}##}$\hfil\tabskip\centering
    &\llap{$\displayt@gpar##$}\tabskip\z@skip\crcr
    #1\crcr}}

\def\displayt@gpar(#1){(\displayt@g#1 )}

\def\displayt@g#1 {\rm\ifunc@lled{#1}\global\advance\tagnumber by1
        {\def\next{#1}\ifx\next\empty\else\expandafter
        \xdef\csname @qnnum#1\endcsname{\t@ghead\number\tagnumber}\fi}%
  \writenew@qn{#1}\t@ghead\number\tagnumber\else
        {\edef\next{\t@ghead\number\tagnumber}%
        \expandafter\ifx\csname @qnnum#1\endcsname\next\else
        \w@rnwrite{Equation \noexpand\tag{#1} is a duplicate number.}\fi}%
  \csname @qnnum#1\endcsname\fi}

\def\ifunc@lled#1{\expandafter\ifx\csname @qnnum#1\endcsname\relax}

\let\@qnend=\end\gdef\end{\if@qnfile
\immediate\write16{Equation numbers written on []\jobname.EQN.}\fi\@qnend}

\catcode`@=12

\catcode`@=11
\newcount\r@fcount \r@fcount=0
\newcount\r@fcurr
\immediate\newwrite\reffile
\newif\ifr@ffile\r@ffilefalse
\def\w@rnwrite#1{\ifr@ffile\immediate\write\reffile{#1}\fi\message{#1}}

\def\writer@f#1>>{}
\def\referencefile{%			  Stuff to write .REF file
  \r@ffiletrue\immediate\openout\reffile=\jobname.ref%
  \def\writer@f##1>>{\ifr@ffile\immediate\write\reffile%
    {\noexpand\refis{##1} = \csname r@fnum##1\endcsname = %
     \expandafter\expandafter\expandafter\strip@t\expandafter%
     \meaning\csname r@ftext\csname r@fnum##1\endcsname\endcsname}\fi}%
  \def\strip@t##1>>{}}

\def\citeall#1{\xdef#1##1{#1{\noexpand\cite{##1}}}}
\def\cite#1{\each@rg\citer@nge{#1}}	% Variable No. of args, separated by

\def\each@rg#1#2{{\let\thecsname=#1\expandafter\first@rg#2,\end,}}
\def\first@rg#1,{\thecsname{#1}\apply@rg}	% each@ag is a general purpose
\def\apply@rg#1,{\ifx\end#1\let\next=\relax%	  variable no. of arg. macro.
\else,\thecsname{#1}\let\next=\apply@rg\fi\next}% args separated by commas

\def\citer@nge#1{\citedor@nge#1-\end-}	% Check for M-N range (M and N numbers)
\def\citer@ngeat#1\end-{#1}
\def\citedor@nge#1-#2-{\ifx\end#2\r@featspace#1 % Single argument
  \else\citel@@p{#1}{#2}\citer@ngeat\fi}	% M-N range of arguments
\def\citel@@p#1#2{\ifnum#1>#2{\errmessage{Reference range #1-#2\space is bad.}%
    \errhelp{If you cite a series of references by the notation M-N, then M and
    N must be integers, and N must be greater than or equal to M.}}\else%
 {\count0=#1\count1=#2\advance\count1 by1\relax\expandafter\r@fcite\the\count0,
  \loop\advance\count0 by1\relax%	  Loop from M to N
    \ifnum\count0<\count1,\expandafter\r@fcite\the\count0,%
  \repeat}\fi}

\def\r@featspace#1#2 {\r@fcite#1#2,}	% Eat spaces at beginning or end of arg
\def\r@fcite#1,{\ifuncit@d{#1}%		  Cite individual reference
    \newr@f{#1}%
    \expandafter\gdef\csname r@ftext\number\r@fcount\endcsname%
                     {\message{Reference #1 to be supplied.}%
                      \writer@f#1>>#1 to be supplied.\par}%
 \fi%
 \csname r@fnum#1\endcsname}
\def\ifuncit@d#1{\expandafter\ifx\csname r@fnum#1\endcsname\relax}%
\def\newr@f#1{\global\advance\r@fcount by1%
    \expandafter\xdef\csname r@fnum#1\endcsname{\number\r@fcount}}

\let\r@fis=\refis			% Save old \refis, redefine
\def\refis#1#2#3\par{\ifuncit@d{#1}%      Use two params #2 #3 to strip blank
   \newr@f{#1}%
   \w@rnwrite{Reference #1=\number\r@fcount\space is not cited up to now.}\fi%
  \expandafter\gdef\csname r@ftext\csname r@fnum#1\endcsname\endcsname%
  {\writer@f#1>>#2#3\par}}

\def\ignoreuncited{%   redefine \refis if ignoring uncited references
   \def\refis##1##2##3\par{\ifuncit@d{##1}%
    \else\expandafter\gdef\csname r@ftext\csname r@fnum##1\endcsname\endcsname%
     {\writer@f##1>>##2##3\par}\fi}}

\def\r@ferr{\endreferences\errmessage{I was expecting to see
\noexpand\endreferences before now;  I have inserted it here.}}
\let\r@ferences=\references
\def\references{\r@ferences\def\endmode{\r@ferr\par\endgroup}}

\let\endr@ferences=\endreferences
\def\endreferences{\r@fcurr=0%		  Save old \endreferences, redefine
  {\loop\ifnum\r@fcurr<\r@fcount%	  Loop over refnum and produce text
    \advance\r@fcurr by 1\relax\expandafter\r@fis\expandafter{\number\r@fcurr}%
    \csname r@ftext\number\r@fcurr\endcsname%
  \repeat}\gdef\r@ferr{}\endr@ferences}

% Save old \endpaper, redefine it to write parting message.

\let\r@fend=\endpaper\gdef\endpaper{\ifr@ffile
\immediate\write16{Cross References written on []\jobname.REF.}\fi\r@fend}

\catcode`@=12

\citeall\refto		% These macros will generate citations
\citeall\ref		%
\citeall\Ref		%

\ignoreuncited
\singlespace

\title{WHY WE DON'T NEED QUANTUM PLANETARY DYNAMICS: 
DECOHERENCE AND THE CORRESPONDENCE PRINCIPLE FOR CHAOTIC SYSTEMS.}
\bigskip
\author{Wojciech Hubert Zurek$^{(1)}$ and Juan Pablo Paz$^{(1,2)}$}
\affil{$(1)$: Theoretical Astrophysics, Los Alamos National Laboratory, 
Los Alamos, NM 87545, USA}
\affil{$(2)$: Departamento de F\'\i sica, Facultad de Ciencias Exactas y 
Naturales, Pabell\'on 1, Ciudad Universitaria, 1428 Buenos Aires, 
Argentina}
\abstract{ Violation of correspondence principle may occur for very 
macroscopic but isolated quantum systems on rather short timescales 
as illustrated by the case of Hyperion, the chaotically tumbling moon 
of Saturn, for which quantum and classical predictions
are expected to diverge on a timescale of approximately 20 years. 
Motivated by Hyperion, we review salient features of ``quantum chaos'' and show
that decoherence is the essential ingredient of the classical limit, as it 
enables one to solve the apparent paradox caused by the breakdown 
of the correspondence principle for classically chaotic systems. }
\oneandahalfspace

\subhead{1. Introduction}

Is the correspondence principle valid for quantum systems whose classical
counterparts are chaotic? This question has been at the center of a debate that 
has taken place in recent years within the community of scientists interested
in {\it quantum chaos} \refto{QuantumCh, Berry, Ford}. In this 
paper we will argue that the 
apparent failure of the correspondence principle is cured by decoherence, 
which is an essential ingredient  to  properly define a
classical limit. We shall begin by schematically presenting the problem.
Subsequently, we shall sketch the solution provided by decoherence. 

There is no unique way to state the correspondence principle. Indeed, 
various approaches can be found in the literature. All of them predict
failure of the quantum-classical correspondence when applied to quantum systems 
which are classically chaotic. What most authors seem to understand by 
correspondence is the rough idea that quantum mechanics, when applied 
to macroscopic systems must agree with the predictions of classical 
Newtonian dynamics. For Bohr and Heisenberg -- and most quantum mechanics 
textbooks -- the correspondence principle is expected to be valid
in the limit of large quantum numbers, $\hbar\rightarrow 0$, $1/n$ or the like. 
Another way of looking at this issue, based on Ehrenfest theorem, is to note
that for a sharply peaked wave packet, characterized by large 
occupation numbers, the expectation values $<x>$ and $<p>$ follow
classical trajectories satisfying Newton's laws. 

As mentioned above, 
in any of its forms, correspondence principle seems to be in trouble when 
applied to systems which are classically chaotic. To clearly state 
the problem \refto{ZPPRL} it is convenient to use the phase 
space formulation of quantum mechanics based on the Wigner function $W(x,p)$ 
whose evolution equation (entirely equivalent to Schr\"odinger equation)
reads \refto{Wigner}:
$$
\eqalign{
\dot W=&\{H,W\}_{MB}\cr
=&
\{H,W\}_{PB}+\sum_n{\hbar^{2n}(-1)^n\over(2n+1)!2^{2n}}\partial^{(2n+1)}_x V
\partial^{(2n+1)}_pW.\cr}\eqno(wignereq)
$$
The operator in the right hand side of (1) is known as the Moyal bracket. 
When the potential $V$ is analytic, Moyal bracket can be expanded to yield 
Liouville equation with quantum corrections, as it is illustrated above.
The first term in that expansion is the ordinary Poisson bracket, 
which generates the Liouville flow in the phase space according 
to which a classical distribution function evolves. 
The sum in the second term contains all the quantum mechanical effects. 
Therefore, Liouville flow in phase space (and consequently, classical 
dynamics) is obtained from the basic quantum picture as long as the 
quantum corrections appearing in (1) are negligible. 

Consider now an initial state that 
corresponds to a Gaussian packet 
which is round and smooth over scales much larger than $\hbar$ (i.e, 
$\Delta x_0\Delta p_0\gg\hbar$). 
For such a state the sum in (1) is negligible 
since it involves derivatives of a smooth function. Indeed one can see 
that the n--th order term in the sum is proportional 
to $\bigl(\hbar/\chi\sigma_p\bigl)^{2n}$ where $\sigma_p$ is the 
scale over which the Wigner function varies along the momentum direction
and $\chi$ is the scale over which the potential is nonlinear 
(e. g., $\chi \simeq \sqrt{\partial_x V \over \partial_{xxx} V}$) 
within the range where
it is influencing the evolution of the state. 
Therefore, a smooth initial state will start evolving
with negligible quantum corrections. Each point in phase space will start
following its corresponding classical trajectory. However, this state of 
affairs cannot last forever: After some time $t_{\hbar}$, 
the Wigner function that evolves according to equation \(wignereq) 
will start looking different from a classical distribution function 
which has originated from the same initial condition but which has 
evolved according to Liouville equation. 
>From that time, the difference between the 
quantum expectation values $<x^k(t)>, 
<p^k(t)>$, calculated from the Wigner function, and their classical 
counterparts obtained from the classical distribution function  
will tend to increase.  

To see if this obvious property of quantum evolution poses a problem for
the correspondence principle, the relevant question is: ``How long is 
the {\it correspondence breakdown time} $t_{\hbar}$?''. 
The answer to this question is dramatically different depending on the
nature of the evolution -- that is, on whether the system is 
classically chaotic or integrable. For a classically chaotic system, 
an initially smooth phase space patch will be exponentially stretched in 
the directions corresponding to positive Lyapunov exponents. As the volume
in phase space is preserved by the Liouville flow, $W(x,p)$, will tend to 
shrink in other directions. Consequently, derivatives of the Wigner function 
will grow exponentially fast generating the growth of the 
``quantum corrections''. The time after which the initially small 
quantum corrections become comparable with the Liouville term 
is \refto{ZPPRL, ZPPhysica}:
$$
t_{\hbar} \simeq {1\over\lambda}\ln({\chi\sigma_p\over\hbar}) \ , \eqno(tchi)
$$
where $\lambda$ is the Lyapunov exponent while $\chi$ and $\sigma_p$ are
defined above. A similar estimate, 
$$
t_r \simeq {1 \over \lambda} \ln ({ A_0 \over \hbar}) \ , \eqno(tBZ)
$$
was obtained earlier on the basis of a rather different argument by 
Berman and Zaslavsky \refto{BermanZas}. Above, $A_0$ is some characteristic
action which -- for macroscopic systems -- is presumably very large compared
with the Planck constant.
Moreover, typical $A_0$ is large (and often very large) compared with the
volume in the phase space $\chi \sigma_p$ associated with the initial 
conditions. Thus, $t_r \geq t_{\hbar}$ is likely to be satisfied.

By contrast, 
for integrable systems, analogous correspondence breakdown occurs only at;
$$
t_{\hbar}^{(int)} \simeq {1\over\Omega}\bigl({A_0\over\hbar})^\alpha \ , 
\eqno(tbreakint) 
$$
where $\Omega$ is some dynamical frequency, $A_0$ is a characteristic
action (that plays the role of the product $\chi\sigma_p$ in (2)) and $\alpha$ 
is some positive power. The difference between the behavior displayed 
in equations \(tchi) and \(tBZ) on the one hand and \(tbreakint) on the
other is quite dramatic: Quantized counterparts of classically chaotic systems 
depart from classical behavior much sooner than classically integrable systems 
-- on an uncomfortably short timescale $t_{\hbar}$ which increases only
logarithmically with the decrease of the Planck constant. 

\subhead{2. For how long will Hyperion be classical?}

After taking a superficial look at equations (2), (3), and (4) one may be 
tempted to conclude that there is no problem at all with the correspondence 
principle: Taking the $\hbar\rightarrow 0$ limit in both equations one obtains 
$t_r\rightarrow \infty$. However, this is not enough. 
Thus, classicality simply does not follow ``as $\hbar \rightarrow 0$'' in most 
{\it physically} interesting cases (including chaos). Planck constant is 
$\hbar=1.05459\times10^{-27}$~[erg~s] and -- {\it licentia mathematica} to vary 
it notwithstanding -- it is a {\it constant}. The right question is: 
``What is the value of $t_r$ (or $t{\hbar}$ for macroscopic quantum systems?''. 
And this is precisely  where the true problem with the correspondence 
principle shows up since one easily discovers that (2) is simply too short, 
even for systems  where classical behavior is expected and observed.

A particularly remarkable example we have found 
is provided by Hyperion, one of the moons of Saturn. 
Hyperion is a highly aspherical object whose principal radii measure
$(150\times 145\times 114\pm 10)$[km] (see \refto{Wisdom}). 
Its irregular motion has been originally detected by monitoring changes in 
its luminosity and has been tracked by the recent observations carried out
during the Voyager 2 mission: Hyperion is tumbling in a chaotic regime 
while orbiting around Saturn. 
The Lyapunov exponent that characterizes this chaotic motion, while not 
directly measured, it is believed to be of the order of two orbital periods, 
which are $21$ days. To estimate correspondence breakdown time $t_r$ 
we should find out the action $A_0$ or the value of 
the product $\chi\sigma_p$. A generous overestimate 
of the $A_0$ is given by the product of Hyperion's 
orbital kinetic energy (which is certainly larger than the energy associated 
with its tumbling motion) and its $21$--day period. 
This yields $t_r \approx 100/\lambda \approx 20$~[yrs]. 
Therefore, given that $t_r$ is obviously orders of magnitude less than 
Hyperion's age one would expect the moon to be in a very non-classical 
superposition, behaving in a flagrantly quantum manner. 
In particular, after a time of this order the phase angle characterizing the
orientation of Hyperion should become coherently spread over macroscopically
distinguishable orientations -- the wavefunction would be a coherent 
superposition over at least a radian. 
This is certainly not the case, 
Hyperion's state and its evolution seem perfectly classical.
Why? The answer (which we outlined in our paper \refto{ZPPRL}, 
as well as elsewhere \refto{ZPPhysica, ZPreply}) is provided by decoherence.

\subhead{3. Decoherence and Classicality}

The interest in the process of decoherence did not arise in the field of 
quantum chaos. Its importance and the role of {\it environment induced 
superselection} 
has been first recognized in the context of quantum measurement theory
\refto{Decoherence, JoosZeh, zurekpt,zurek93,ZurekPT}. As 
we will see, the reason why decoherence can solve the ``correspondence
paradox'' is basically the 
same that makes it an essential ingredient to explain the transition 
from  quantum to classical in other contexts. 

Decoherence is the process of loss of (phase) coherence by the system caused
by the interaction with the external or internal degrees of freedom which
cannot be followed by the observer and are summarily called `the environment'. 
Different states in the Hilbert space of the system of interest show 
various degrees of susceptibility to decoherence. States which are least
susceptible (i.e., take longest to decohere) form the {\it preferred basis} 
(also known as the {\it pointer basis} in the context of quantum 
measurement)\refto{Decoherence,zurekpt,zurek93,ZurekPT}. 
Preferred states are singled out by the interaction between the system and 
the environment. In this way, an {\it environment induced superselection rules}
arise, which effectively outlaw arbitrary
superpositions. Thus, even though the superposition principle is valid in 
a closed quantum system, it is invalidated by decoherence for systems 
interacting with their environments. 

All of the macroscopic quantum systems 
we encounter in our everyday existence, as well as our own memory
and information processing hardware (e.g., neurons, etc.) are macroscopic 
enough and sufficiently strongly coupled to the environment to be 
susceptible to decoherence, which will eliminate truly quantum superpositions
on a very short timescale. This process is absolutely essential in the 
transition from quantum to classical in the context of quantum measurements
(where the classical apparatus tends to be very macroscopic) although 
resolutions based on decoherence may not be easily palatable to everyone
(i.e., see comments on decoherence in the April 1993 issue of {\it Physics 
Today} and also \refto{CasatChir}). 

The timescale on which decoherence takes place can be estimated by solving 
a specific example: a one dimensional particle moving in a potential 
$V(x)$ coupled through its position with a thermal environment -- e.g. 
with a collection of harmonic oscillators at a temperature $T$ \refto{QBM}. 
Under the appropriate assumptions (Markovian regime) one can derive 
the following 
equation for the reduced Wigner function of the preferred particle:
$$
\dot W=\{H,W\}_{MB}+2\gamma\partial_pp W + D\partial^2_{pp}W\eqno(wignereq2)
$$
The last two terms in this equation carry all the effects of the 
environment producing (respectively) 
relaxation and diffusion. $D=2 m\gamma k_BT$ is 
the diffusion coefficient and $\gamma$ the relaxation rate. The diffusion 
term is the one responsible for decoherence: Consider  
the Wigner
function corresponding to a superposition of two localized states separated
by a distance $\Delta x$. This function is the sum of three terms, two 
direct contributions and an interference term. The interference term 
is modulated by an 
oscillatory function of the form $\cos(p\Delta x/\hbar)$. Thus, 
when evolving under equation \(wignereq2) these 
``interference fringes'' tend to be exponentially damped by the 
decoherence term (which, as we mentioned, is the last one in \(wignereq2) 
and leaves the direct terms essentially uneffected). 
The exponential decay of the interference takes place 
in a decoherence timescale\refto{Zurek86};
$$
\tau_D=\gamma^{-1} {\hbar^2\over D (\Delta x)^2}= \tau_R 
({\lambda_{dB}\over\Delta x})^2 \ , \eqno(dectime)
$$
where $\lambda_{dB}=(\hbar^2/2mk_BT)^{1/2}$ is the thermal de Broglie 
wavelength and $\tau_R=\gamma^{-1}$ is the relaxation timescale.  

Two remarks are in order: (i) The decoherence timescale $\tau_D$ is much
shorter than the relaxation timescale $\tau_R$ for all macroscopic 
situations, as typical thermal de Broglie wavelengths of macroscopic 
bodies are many orders of magnitude smaller than macroscopic separations
$\Delta x$. (ii) The devastating effect of decoherence on superpositions 
of position can be traced back to the preferential monitoring of that 
observable ($x$) by the environment, which was coupled to the position of
the system of interest. This also tends to be the case in general: Interaction
potentials depend on position and, therefore, allow the environment to 
monitor $x$\refto{Decoherence,zurekpt,phz}. 

As a result of the action of
the decoherence term, the vast majority of states which could in principle 
describe the system of interest would be, in practice, eliminated by 
the resulting environment - induced superselection. Only localized
states will be able to survive. They will form a preferred basis. For 
they will be much more stable than their coherent superpositions
(even though they will be in general still somewhat  
unstable under the joint action of the self--hamiltonian and the 
environment). For example, 
in an underdamped harmonic oscillator the preferred states turn out to be
the familiar coherent states\refto{zhp}: Oscillator dynamics rotates all of 
the states, which, in effect, translates spread in position into spread 
in momentum (and vice versa) every quarter period of the oscillation. 
As a result, coupling to position can be quite faithfully represented in the 
``rotating wave approximation'' which makes the master equation symmetric in 
$x$ and $p$ \refto{Louisell}. 
Hence, coherent states will minimize entropy production and are therefore
selected by {\it predictability sieve} as classical \refto{zurek93, ZurekPT}. 
By contrast, 
for superpositions of coherent states entropy production will happen on a very
much shorter decoherence timescale.

Summarizing, {\it environment induced decoherence} is a natural process
that prevents the stable existence of generic quantum states which are 
spread over a large region of phase space. At this point, one may discover 
that this is precisely what we need to recover the correspondence principle for 
classically chaotic systems. Indeed, chaotic dynamics is especially effective
in transforming a smooth initial state into a highly delocalized one with 
a complicated Wigner function and a lot of small scale structure. Decoherence 
will naturally compete against this process trying to favor smooth 
and localized states, or mixtures thereof. The result of this competition is 
a very interesting balance which enables us to recover 
the correspondence principle. 

\subhead{4. Decoherence, exponential instability and correspondence.}

To understand the nature of the compromise between decoherence and 
exponential instability it is worth studying this process under simplifying
assumptions \refto{ZPPRL}. We will be interested in the regime in which 
the coupling to the 
environment is sufficiently weak so that the damping (represented by the second 
term in \(wignereq2)) is negligible. This is the so--called ``reversible 
classical limit''\refto{zurekpt,Zurek86,phz} which in integrable systems yields 
reversible classical trajectories for localized (i. e. gaussian) states 
but still eliminates non--local superpositions (this limit is achieved 
by letting $\gamma$ approach zero but keeping $D$ constant so that decoherence 
continues to be effective). In this limit, equation
\(wignereq2) can be rewritten as:
$$
\dot W=\{H,W\}_{MB}+ D\partial^2_{pp}W.\eqno(wignereq3)
$$

Let us consider, as we did above, an initial state which is smooth. Thus, 
the Wigner function initially evolves under 
the Poisson bracket and the diffusion term.
Then, in the neighbourhood of any point, equation \(wignereq2) can be easily 
expanded along the unstable ($\lambda_i^+>0$) and stable ($\lambda_i^-<0$)
directions in phase space ($\sum_i(\lambda_i^-+\lambda_i^+)=0$). 
Diffusion will have little influence on the evolution of $W$ along the 
unstable directions: 
$W$ will be stretched simply as a result of the dynamics, so 
that the gradients along these directions will tend to decay anyway, without
assistance from diffusion. By contrast, squeezing which occurs along the 
contracting directions will tend to be opposed by the diffusion. This will 
lead to a steady state with the solution asymptotically approaching a Gaussian 
with a half--width given by the critical dispersion:
$$
\sigma_{c_i}^2=2D_i/|\lambda_i^-|\eqno(sigmacrit)
$$
where $\lambda_i^-$ is the (negative) Lyapunov exponent along the stable 
direction and $D_i$ is the diffusion coefficient along the same direction. 
Below, we will assume that the diffusion is isotropic (as would be the 
case in the rotating wave approximation). Thus, after some time (and in 
the absence of folding -- the other aspect of chaos which we will discuss 
below) the Wigner function will evolve into a multidimensional 
``hyper--pancake,'' still stretching along the unstable directions but 
with its width limited from below in the stable directions by equation
\(sigmacrit). 

The existence of this critical width, an important consequence of the 
interplay between decoherence and exponential instability, has remarkable 
consequences concerning the rate of entropy production. In fact, 
at this stage, entropy will be approximated by the logarithm of the 
effective volume of the hyper--pancake. As its extent 
in the stable direction is fixed by the critical width \(sigmacrit), its
volume will tend to increase at a rate given by the positive exponents. 
Consequently, 
$$
\dot H \approx \sum_i\lambda_i^+.\eqno(hdot)
$$
This constant rate will set in after a time larger than the decoherence
timescale  
and after a time over which the initial Wigner distribution becomes squeezeed
by the dynamics to the dimension of order of the critical dispersion 
$\sigma_{c_i}$. Equation \(hdot) will be valid until the pancake fills in 
the available phase space and the system reaches (approximately) uniform 
distribution over the accessible part of the phase space, that is after 
a time defined by;
$$
t_{eq}=(H_{eq}/H_0)/\dot H,\eqno(eqtime)
$$
where $H_0$ is the initial entropy, and $H_{eq}$ is the entropy uniformized
by the chaotic dynamics. 

Astute reader will note that $H_{eq}$ above need not be a true equilibrium 
entropy with the temperature given by $T$. Rather, it will correspond to 
dynamical quasi--equilibrium -- the approximately uniform distribution over 
this part of the phase space which (given specific initial conditions)
is accessible to the chaotic system as a result of its dynamics.
The corresponding timescale will have a similar dependence on
$\hbar$ as the timescale $t_{\hbar}$ defined by \(tchi). This is because entropy
is approximately given by the logarithm of the volume of the phase space over 
which the probability distribution has spread in the units of Planck constant.
Nevertheless, $t_{\hbar}$ (or $t_r$) and $t_{eq}$ depend on rather different 
aspects of the initial and final state, and one can expect the correspondence 
breakdown time to be typically a fraction of $t_{eq}$.

The existence of the critical width \(sigmacrit) is a property of 
classically chaotic systems. By contrast, in integrable systems stretching of 
the corresponding hyper--pancake in 
phase space will proceed only polynomially. Thus, even when it will get to 
the stage at which, in the contracting direction, diffusion will become
important, stretching in the unstable direction will be only polynomial
(rather than exponential). Consequently, the volume of the hyper--pancake
will increase only as some power of time. Hence, the entropy will grow
only logarithmically as the entropy production rate will fall as
$\dot H\propto 1/t$: It will take exponentially long to approach dynamical
quasi--equilibrium. 
This difference in behavior between chaotic and integrable {\it open} 
quantum systems is striking and can be used as a defining feature of
quantum chaos \refto{ZPPhysica}. 

Let us now focus on the recovery of the correspondence principle. 
Decoherence limits the extent over which the wavefunction can remain 
coherent. This is because a finite minimal dispersion in momentum 
\(sigmacrit) corresponds to quantum coherence over distances no longer than:
$$
l=\hbar/\sigma_c=\hbar(2D/\lambda)^{-1/2}.\eqno(lcrit)
$$
Thus, when the scale $\chi$ on which nonlinearities in the potential are significant is small compared to the extent of the wavefunction
$$
\chi\ll l\eqno(chivsl)  
$$
decoherence will have essentially no effect. Evolution will remain purely
quantum and will be generated by the full Moyal bracket. 

By contrast, when the opposite is true, the evolution will never squeeze 
Wigner distribution function enough for the full Moyal bracket to be 
relevant. Poisson bracket will suffice to approximate the flow of probability
in phase space. The inequality characterizing this case can be written in 
a manner reminiscent of the Heisenberg indeterminacy principle:
$$
\hbar\ll \chi\sigma_c.\eqno(ineqclass)
$$
That is, as long as decoherence keeps the state vector from becoming too
narrow in momentum, Poisson bracket is all that is required to evolve 
the Wigner function. Therefore, inequality \(ineqclass) defines the 
regime in which one recovers the correspondence principle.  

There is one more interesting regime where the chaotic motion is 
dynamically reversible (that is, $\dot H=0$) even if the system satisfies
inequality \(ineqclass). This happens when the initial patch in phase 
space is large (volume much larger than the Planck volume -- initial 
entropy larger than a single bit) and regular. Then the initial stage
of the evolution will proceed reversibly, in accord with the Poisson 
bracket generated flow. Decoherence will have little effect. This is 
because its influence will set in only as the dimension of the Wigner
distribution in the contracting direction will approach the critical 
dispersion $\sigma_c$: In a simple example (see \refto{ZPPRL})
the entropy production will increase as:
$$
\dot H = \lambda{1\over \Bigl(1+({\sigma_p^2(0)\over\sigma_c^2}-1)\exp(
-2\lambda t)\Bigr)}\eqno(hdotapp)$$

So far, we have not taken into account (or, at least, not taken into 
account explicitely) the other major characteristic of chaos: In addition 
to exponential instability, chaotic systems ``fold'' the phase space
distribution. While this problem may require further study, we believe that
the fundamentals of folding are already implicit in the above discussion: 
Folding will happen on the scale $\chi$ of nonlinearities in the 
potential (which will typically -- but not always -- coincide with the size of 
the system, as it is defined by the range of its classical trajectory). 
Hence, preventing the system from maintaining coherence over distances
of the order of $\chi$ will also ascertain its classical behavior in course
of folding. There will simply be no coherence left between the fragments
of the wavepacket which will come into proximity as a result of folding, if
they had to be separated by distances larger than $l$ in the course of the 
preceding evolution. Thus, folding will proceed as if the system was
classical, but with a proviso: After sufficiently many folds the distribution
function (which in the stable direction cannot shrink to less than $\sigma_c$) 
will simply fill in the available phase space. This will be achieved in 
the previously defined equilibrium timescale $t_{eq}$. These conclusions are
consistent with the studies of quantum maps corresponding to open 
quantum systems such as the ``standard map'' carried out by Graham and 
his coworkers\refto{Graham}.

\subhead{5. Summary}

We have argued that decoherence is the essential ingredient that 
enables us to solve the apparent paradox caused by the lack of validity
of the correspondence principle for classically chaotic systems. 
Violation of correspondence principle may occur for isolated quantum
systems on a rather short timescales as illustrated by Hyperion, the 
chaotically tumbling moon of Saturn. Decoherence or, more precisely, 
the continuous monitoring by the environmental degrees of freedom and 
the ensuing ``reduction'' of the quantum state of Hyperion 
(or any other open quantum system) -- continually
forces them to be classical. This  
process in turn leads to {\it environment - induced superselection} 
as a result of which 
only a small subset of preferred {\it pointer states} in the 
Hilbert space of the system are sufficiently immune to be 
predictable and to belong to ``classical reality''. 

Decoherence gurarantees the validity of the correspondence
principle by precluding the growth of gradients of the Wigner function
ensuring that the quantum corrections to equation \(wignereq2) remain small.
This process is accompanied by the increase of entropy: 
The information
acquired by the environment is lost to the observer. 
We also explained why entropy production is so different for quantum 
open systems which are classically regular or chaotic: In the last case, the 
exponential instability tends to create fine structure in the Wigner 
function $W$ but this process is stopped by the diffusion induced by the 
environment. Thus, $W$ cannot squeeze beyond the critical width 
$\sigma_c$ given by \(sigmacrit). At this point entropy starts growing linearly 
in time at a rate fixed by the Lyapunov exponent. 
This is how most of the entropy in 
an open chaotic system starting from a low entropy, localized ($\sim$~classical)
state will be produced. Eventually, close to equilibrium the support of $W$ will
fill in the phase space available to the system at the energy shell consistent 
with the initial conditions, and the entropy production rate will decrease
to halt at $H_{eq}$. This will occur near $t_{eq}\simeq\lambda^{-1}H_{eq}/H(0)$,
where $t_{eq}$ is the timescale for reaching equilibrium. By contrast, in 
a regular (integrable) system trajectories diverge (or become squeezed) only 
with a power of time. Hence, the support of $W$ in presence of diffusion will
increase only as $t^n$, so that nearly all of the entropy is gained very slowly,
while $\dot H \sim 1/t$. While we have argued for these conclusions with 
the help of an exactly solvable model -- unstable oscillator (which is of course not chaotic, but 
represents well 
the local instability of chaotic evolution) -- we believe that our conclusions 
concerning $\dot H$ will hold for $t_{\hbar} < t < t_{eq}$ for chaotic systems. 
Indeed, we have conjectured that entropy production rate in a slightly open 
system may be a good ``diagnostic'' to distinguish between chaotic and regular 
quantum systems \refto{ZPPhysica}.

Decoherence caused by the environment (considered unsatisfactory by some
authors \refto{CasatChir})
is not a subterfuge of a theorist, but a fact of life: 
Macroscopic systems are exceedingly difficult to isolate from their environments
for a time comparable to their dynamical timescale. Moreover, even if their 
energy is almost perfectly conserved, purity of their wavepacket may not be 
assured: As the examples studied in our paper and elsewhere indicate, the 
boundary between the system and the environment may be nearly impenetrable to 
energy, but very ``leaky'' for information. This imperfect isolation is, we 
believe, the reason why classical behavior emerges from the quantum substrate.

Sections 3-5 of this manuscript are based in part on the paper which was 
also presented at a meeting on {\it Quantum Complexity in Mesoscopic Systems},
and will appear in its proceedings \refto{ZPPhysica}.

\references

\refis{CasatChir} G. Casati and B. V. Chirikov, Comment on {\it Decoherence, 
chaos and the second law}, Phys. Rev. Lett. (1995) to appear. 

\refis{ZPreply} W. H. Zurek and J. P. Paz, Reply to Casati and Chirikov, 
Phys. Rev. Lett. (1995) to appear. 

\refis{BermanZas} G. P. Berman and G. M. Zaslavsky, Physica {\bf A91}, 450 
(1978).

\refis{Berry} see 
M. B. Berry, ``Introductory remarks'' in {\it Adriatico Research 
Conference and Miniworkshop on Quantum Chaos}, edited by H. A. Cerdeira et al 
(Word Scientific, SIngapore, 1991). 

\refis{Wisdom} J. Wisdom, Icarus {\bf 72}, 241, (1987).

\refis{6.} W. H. Zurek, Phys. Rev. {\bf D24}, 1516 (1981); {\it ibid} 
{\bf 26}, 1862 (1982); Physics Today {\bf 44}, 36 (1991); J. P. Paz, S. Habib
and W. H. Zurek, Phys. Rev. {\bf D47}. 488 (1993).

\refis{ZPPhysica} W. H. Zurek and J. P. Paz, Physica {\bf D} to appear (1995).

\refis{WheelerZurek} J. A. Wheeler and W. H. Zurek, {\it Quantum Theory 
and Measurement}, (Princeton University Press, Princeton, 1983).

\refis{Mazagon} J. J. Halliwell, J. Perez-Mercader, and W. H. Zurek, eds., {\it
Physical Origins of Time Asymmetry}, (Cambridge University Press, 1994). 

\refis{Koln} P. Busch, P. Lahti, and P. Mittelstaedt, eds., {\it Quantum Measurements,
Irreversibility, and the Physics of Information} (World Scientific, Singapore,
1994).

\refis{zurekpt} W. H. Zurek, Physics Today {\bf 44}, 36 (1991).
%; {\it ibid} {\bf 46}, 81 (1993).

\refis{zurek93} W. H. Zurek, Prog. Theor. Phys. {\bf89}, 281 (1993).

\refis{zhp} W. H. Zurek, S. Habib and J.P. Paz, Phys. Rev. Lett. {\bf 70}, 1187 (1993).  

\refis{ZPPRL} W. H. Zurek and J. P. Paz, Phys. Rev. Lett. {\bf 72}, 2508 (1994). 

\refis{phz} J. P. Paz, S. Habib and W. H. Zurek,  Phys. Rev. {\bf D 47} (1993) 488.

\refis{qbm} B. L. Hu, J. P. Paz and Y. Zhang, Phys. Rev. {\bf D 45}, 
(1992) 2843; {\it ibid} {\bf D 47} (1993) 1576.

\refis{Chirikov} B.V. Chirikov, Phys. Rep. {\bf 52}, 263 (1979); G. M. Zaslavsky, Phys. Rep. 
{\bf 80}, 157 (1981); M--J. Giannoni, A. Voros, J. Zinn--Justin, eds., {\it Chaos and Quantum 
Physics}, Les Houches Lectures LII (North Holland, Amsterdam, 1991). 

\refis{Ford} J. Ford and G. Mantica, Am. J. Phys. {\bf 60}, 1086 (1992).

\refis{Hamiltonian} See, e.g. G. Ioos, R. H. G. Hellmann and R. Stora eds., 
{\it Chaotic Behavior in 
Deterministic Systems}, Les Houches Lectures XXXVI (North Holland, Amsterdam, 1983); 
R. S. Mc Kay and J. O. Meiss eds., {\it Hamiltonian Dynamical Systems} (Hilger, 
Philadelphia, 1987).

\refis{QuantumCh} M. B. Berry, Proc. Roy. Soc. London, {\bf A413}, 183 
(1987); M. Gutzwiller, 
{\it Chaos in Classical and Quantum Mechanics} (Springer Verlag, New York, 1990); 
F. Haake, {\it Quantum Signature of Chaos} (Springer Verlag, New York, 1990) 
and references therein. 

\refis{Decoherence} W. H. Zurek, Phys. Rev. {\bf D24}, 1516 (1981); {\it ibid}, {\bf D26}, 1862 (1982).

\refis{JoosZeh} E. Joos and H. D. Zeh, Zeits. Phys. {\bf B59}, 229 (1985). 

\refis{ZurekPT} W. H. Zurek, Physics Today, {\bf 46}, 81 (1993).

\refis{PazHZ} J. P. Paz, S. Habib and W. H. Zurek, Phys. Rev. {\bf D47}, 488 (1993).

\refis{ZurekHP} W. H. Zurek, S. Habib and J. P. Paz, Phys. Rev. Lett. {70}, 1187 (1993). 

\refis{Zurek86} W. H. Zurek, pp. 145-149 in {\it Frontiers of Nonequilibrium Statistical Mechanics}, G. 
T. Moore and M. O. Scully eds., (Plenum, New York, 1986). 

\refis{GMH} M. Gell--Mann and J. B. Hartle, Phys. Rev. {\bf D47}, 3345 (1993). 

\refis{QBM} A. O. Caldeira and A. J. Leggett, Physica {\bf 121A}, 587 (1983); W. 
G. Unruh and W. H. Zurek, Phys Rev. {\bf D40}, 1071 (1989); B. L. Hu, J. P. Paz 
and Y. Zhang, Phys. Rev. {\bf D45}, 2843 (1992); {\it ibid} {\bf D47}, 1576 (1993). 

\refis{Louisell} W. H. Louisell, {\it Quantum Statistical Properties of Radiation} (Wiley, New 
York, 1973); 
V. Buzek and P. Knight; {\it Quantum interference, superposition states of light 
and nonclassical effects}, Prog. in Opt. (1993). 

\refis{vonNeumann} J. von Neumann, {\it Mathematical Foundations of Quantum Mechanics}, 
English translation by R. T. Beyer (Princeton Univ. Press, Princeton, 1955); H. D. Zeh, 
{\it The Direction of Time}, (Springer--Verlag, New York, 1991). 

\refis{Blatt} J. M. Blatt expressed a similar suspicion in a compelling, although less rigorous 
and completely classical paper, Progr. Theor. Phys. {\bf 22}, 745 (1959)

\refis{Milburn} G.J. Milburn and C.A. Holmes, Phys. Rev. Lett. {\bf 56}, 2237 (1986).

\refis{Casati} G. P. Berman and G. M. Zaslawsky, Physica {\bf A91}, 450 (1978);
G. Casati, B. V. Chirikov, F. M. Izraileev and J. Ford, {\it Lectures Notes in 
Physics} {\bf 93}, (Springer--Verlag, New York, 1979). 

\refis{Graham} T. Dittrich and R. Graham, Z. f\"ur Phys. {\bf B62}, 
515 (1986); T. Dittrich and R. Graham, Ann. Phys. (NY) {\bf 200}, 363 (1990); 
T. Dittrich and R. Graham, ``Quantum chaos in open systems'', in 
{\it Information Dynamics}, H. Atmanspacher and H. Scheingraber, eds.,
NATO ASI Series, {\bf B256}, 289, (Plenum, 1990) and 
references therein. 

\refis{Wigner} M. Hillery, R.F. O'Connell, M.O. Scully and E. Wigner, Phys. 
Rep. {\bf 106}, 121 (1984).

\endreferences

\end